\documentclass[%
 reprint,
 superscriptaddress,
 amsmath,amssymb,
 aps,
prb,
twocolumn
]{revtex4-2}
\usepackage{dcolumn}
\usepackage{bm}
\usepackage{amsmath}
\usepackage[T1]{fontenc}
\usepackage{lmodern}
\usepackage{xspace}
\usepackage{ulem}
\usepackage{comment}
\usepackage{bbold}
\usepackage{braket}
\usepackage{ascmac}

\ifx\pdfoutput\undefined
\usepackage[dvipdfmx]{graphicx}
\usepackage[dvipdfmx]{hyperref}
\usepackage[dvipdfmx]{color}
\usepackage[dvipdfmx]{xcolor}
\else
\usepackage{graphicx}
\usepackage{hyperref}
\usepackage{color}
\usepackage{xcolor}
\fi

\newcommand{\figref}[2]{\ref{fig:#1}\textbf{#2}}

\newcommand{\hatv}[1]{\hat{\bf #1}}

\newcommand{\QQ}{\mathcal{Q}}
\newcommand{\vp}{\varphi}

\newcommand{\thetaH}{\theta_{\mathrm{Hall}}}

\graphicspath{
{./}
}

\begin{document}

\title{
Current-induced motion of nanoscale magnetic torons \\
over the wide range of the Hall angle
}

\author{Kotaro Shimizu}
\affiliation{RIKEN Center for Emergent Matter Science, Saitama 351-0198, Japan}

\author{Shun Okumura} 
\affiliation{Department of Applied Physics, The University of Tokyo, Tokyo 113-8656, Japan}

\author{Yasuyuki Kato}
\affiliation{Department of Applied Physics, University of Fukui, Fukui 910-8507, Japan}

\author{Yukitoshi Motome}
\affiliation{Department of Applied Physics, The University of Tokyo, Tokyo 113-8656, Japan}

\date{\today}

\begin{abstract}
Current-driven dynamics of spin textures plays a pivotal role in potential applications for electronic devices. 
While two-dimensional magnetic skyrmions with topologically nontrivial spin textures have garnered significant interest, their practical use is hindered by the skyrmion Hall effect --- a transverse motion to the current direction that occurs as a counteraction to the topological Hall effect of electrons by an emergent magnetic field arising from the Berry phase effect. 
Here, we explore current-driven dynamics of three-dimensional topological spin textures known as magnetic torons, composed of layered skyrmions with two singularities called Bloch points at their ends. 
Through extensive numerical simulations, we show that the torons also exhibit a Hall motion, but surprisingly over a wide range spanning from the zero Hall effect, a purely longitudinal motion, to the perfect Hall effect, a purely transverse motion accompanied by no longitudinal motion. 
Such flexible and controllable behaviors stem from anisotropic potential barriers on the discrete lattice, which can be particularly relevant for nanoscale torons recently discovered. 
Our results not only provide an experimental method to probe topology of three-dimensional magnetic textures but also pave the way for future developments in topological spintronics beyond the realm of skyrmions. 
\end{abstract}


\maketitle


\begin{figure*}[tb]
\centering
\includegraphics[width=2.0\columnwidth]{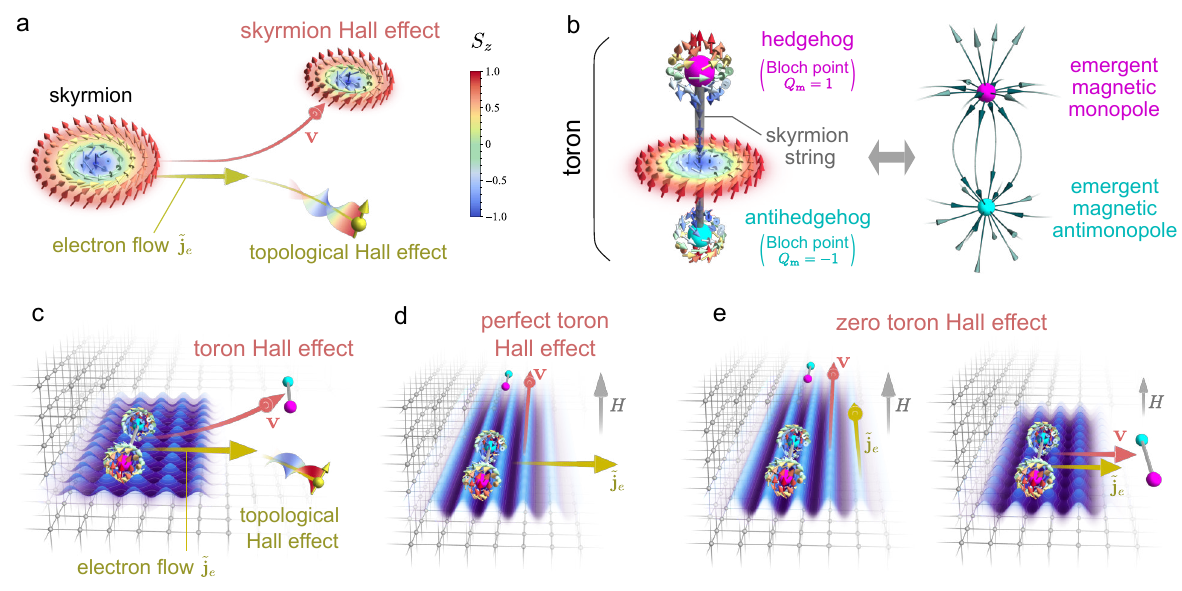}
\caption{
\label{fig:schematic}
\textbf{Schematic illustrations of the topological spin textures and their drift motions driven by an electron flow.} 
{\bf a} Real-space spin configuration of a skyrmion and the skyrmion Hall effect. The small arrows represent the spins and their color displays the $z$ component (see the inset).
Skyrmions show a transverse drift motion as well as a longitudinal one with the velocity $\mathbf{v}$ as a reaction to the topological Hall effect in the electron flow $\tilde{\mathbf{j}}_e$. 
{\bf b} Real-space spin configurations of a magnetic toron, given by a pair of hedgehog (magenta sphere) and antihedgehog (cyan sphere) connected by the skyrmion string (gray line) (left).
Corresponding real-space distribution of the stream of the EMF (right). 
The hedgehog and antihedgehog correspond to a source and sink of the EMF, respectively. 
{\bf c} Toron Hall effect. 
A magnetic toron shows a transverse drift motion similar to the skyrmion Hall effect. 
The blue wavy surface represents the potential for the torons from a discrete lattice structure. 
{\bf d} Perfect toron Hall effect. 
Torons exhibit only a transverse motion without longitudinal one because of strong one-dimensional anisotropy in the lattice potential. 
{\bf e} Zero toron Hall effect. 
Torons exhibit only a longitudinal motion accompanied by no transverse motion. 
The left and right panels correspond to the situations realized in Figs.~\figref{current_velocity}{d},\textbf{e} and \figref{current_velocity}{a}--\textbf{c}, respectively. 
}
\end{figure*}

Research in the field of topological spin textures has opened a new era for electronics and spintronics by leveraging the synergy between electronic and magnetic properties through topology. 
Of particular interest are two-dimensional magnetic skyrmions with swirling spin textures characterized by a topological invariant called the skyrmion number (Fig.~\figref{schematic}{a}). 
The skyrmion was originally hypothesized in the field of particle physics~\cite{Skyrme1961,Skyrme1962}, and later theoretically predicted to be realized in chiral magnets~\cite{Bogdanov1989,Bogdanov1994,Bogdanov1995,Roessler2006} and 
actually observed in experiments in a form of a triangular array~\cite{Muhlbauer2009,Yu2010}. 
Owing to their inherent topological stability, skyrmions have drawn attention as entirely new information carriers in next-generation spintronic devices. 
They also offer a significant advantage in terms of energy consumption, as they can be driven at much lower electric current densities~\cite{Jonietz2010}. 
However, the current-driven skyrmions exhibit a transverse motion relative to the current direction, known as the skyrmion Hall effect~\cite{Jonietz2010,Zang2011,Iwasaki2013,Jiang2017}. 
This is caused by the Magnus force as a counteraction to the topological Hall effect of electrons, as schematically shown in Fig.~\figref{schematic}{a}, arising from the emergent magnetic field (EMF) by the Berry phase effect, $\mathbf{b}(\mathbf{r})$, given by $b_i(\mathbf{r})=\frac{1}{2}\epsilon_{ijk}\mathbf{S}(\mathbf{r})\cdot(\partial_j\mathbf{S}(\mathbf{r})\times\partial_k\mathbf{S}(\mathbf{r}))$, where $\epsilon_{ijk}$ is the Levi-Civita symbol and $\mathbf{S}(\mathbf{r})$ represents the spin at spatial position $\mathbf{r}$~\cite{Berry1984,Volovik1987}. 
This lateral motion presents challenges in device applications of magnetic skyrmions, although various strategies have been proposed, such as utilizing antiferromagnetic skyrmions that do not exhibit the skyrmion Hall effect~\cite{Barker2016,Zhang2016magnetic}.

In three-dimensional space, skyrmions manifest as string-like structures, stacked in the out-of-plane direction. 
This skyrmion string may terminate in some cases, and at these endpoints, unique three-dimensional topological spin textures emerge. 
They are known as the Bloch points and were observed by using 3D tomographic imaging techniques~\cite{Doring1968,Milde2013,Donnelly2017,Hierro2020}. 
These Bloch points are also characterized by a topological invariant $Q_{\mathrm{m}}$ called the monopole charge, which is defined by a surface integral of the EMF $\mathbf{b}(\mathbf{r})$ on a sphere surrounding the Bloch point, and classified into two types depending on the sign of the monopole charge: magnetic hedgehogs and antihedgehogs with positive and negative monopole charges, respectively~\cite{Volovik1987,Kanazawa2016}. 
They typically appear in pairs at both ends of a skyrmion string and such a pair is called the magnetic toron (Fig.~\figref{schematic}{b})~\cite{Leonov2018}. 
Unlike skyrmions, the hedgehogs and antihedgehogs have singularities with vanishing spin length at their cores, which can be regarded as sources and sinks in terms of the EMF, thus acting as emergent magnetic monopoles and antimonopoles, respectively.

Recent experiments have demonstrated that such magnetic torons are stabilized by forming periodic arrangements called the hedgehog lattices (HLs) in chiral magnets MnSi$_{1-x}$Ge$_x$~\cite{Kanazawa2011,Kanazawa2012,Tanigaki2015,Fujishiro2019} and a nonchiral magnet SrFeO$_3$~\cite{Ishiwata2020}. 
Notably, these HLs have remarkably short magnetic pitches, with hedgehogs and antihedgehogs spanning only a few nanometers in size, suggesting a giant EMF around them caused by a strong twist of the spin textures. 
Due to the EMF as well as the topological robustness, Bloch points are expected to exhibit peculiar electromagnetic responses like skyrmions, but their current-driven dynamics remains largely unexplored, except for an isolated Bloch point or toron in the long-wavelength limit~\cite{Hu2021,Lang2023}. 
Given that smaller texture sizes are preferable for EMF-based devices, it is imperative to investigate the current-driven dynamics of these short-pitch HLs to advance future spintronic devices beyond the capabilities of skyrmions.

Here, we examine the responses of nanoscale magnetic torons to an applied electric current by extensive numerical simulations. 
We show that, similar to skyrmions, the torons exhibit a Hall motion, which we call the toron Hall effect (Fig.~\figref{schematic}{c}). 
Strikingly, we discover that the toron Hall effect can be controlled by the electric current and the magnetic field in a wide range including two extremes: a purely transverse motion without any longitudinal one (perfect toron Hall effect; Fig.~\figref{schematic}{d}) and an exclusively longitudinal motion with no transverse one (zero toron Hall effect; Fig.~\figref{schematic}{e}). 
These unique behaviors stem from the modulation of potential barriers on the discrete lattice, which is particularly relevant for the nanoscale torons realized in experiments. 
Furthermore, we reveal that the responses to the current act as efficient electrical probes for topological characteristics hidden behind the spin textures that are challenging to observe experimentally.

\section*{
Toron Hall effect  
\label{sec:MHE}}

\begin{figure*}[tb]
\centering
\includegraphics[width=2.0\columnwidth]{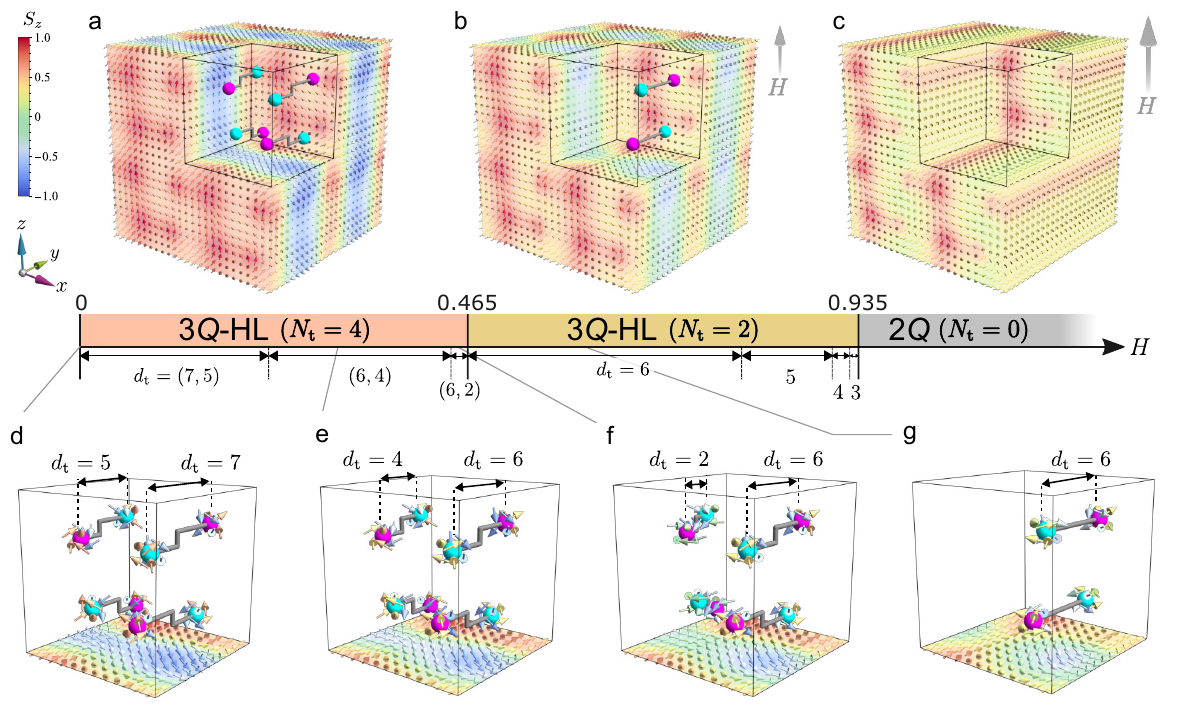}
\caption{
\label{fig:gs}
\textbf{Hedgehog lattice and topological transition in the magnetic field.} 
{\bf a}-{\bf c} Real-space spin configurations and distributions of magnetic torons in the 3$Q$-HL with $N_{\mathrm{t}} = 4$ at $H=0$ (\textbf{a}), the 3$Q$-HL with $N_{\mathrm{t}} = 2$ at $H=0.3$ (\textbf{b}), and the topologically-trivial $2Q$ state with $N_{\mathrm{t}} = 0$ at $H=1.0$ ({\bf c}). 
Here, $N_{\mathrm{t}}$ denotes the total number of the torons in the MUC represented by the black cubes. 
The graphical notations are common to those in Fig.~\ref{fig:schematic}. 
The lower bar displays the phase diagram while increasing the magnetic field $H$. 
{\bf d}-{\bf g} Real-space distributions of the magnetic torons within the MUC for the ground state at $H=0$ (\textbf{d}), $H=0.3$ (\textbf{e}), $H=0.45$ (\textbf{f}), and $H=0.6$ (\textbf{g}). 
Here, $d_{\mathrm{t}}$ denotes the length of the skyrmion string in the $y$ direction. 
As $H$ increases, the hedgehogs and antihedgehogs approach each other along the skyrmion strings (\textbf{d}-\textbf{f}), and half of them disappear with pair annihilation at the transition at $H=H_{\rm topo}\simeq 0.465$ (\textbf{f},\textbf{g}). 
}
\end{figure*}

\begin{figure*}[tb]
\centering
\includegraphics[width=2.0\columnwidth]{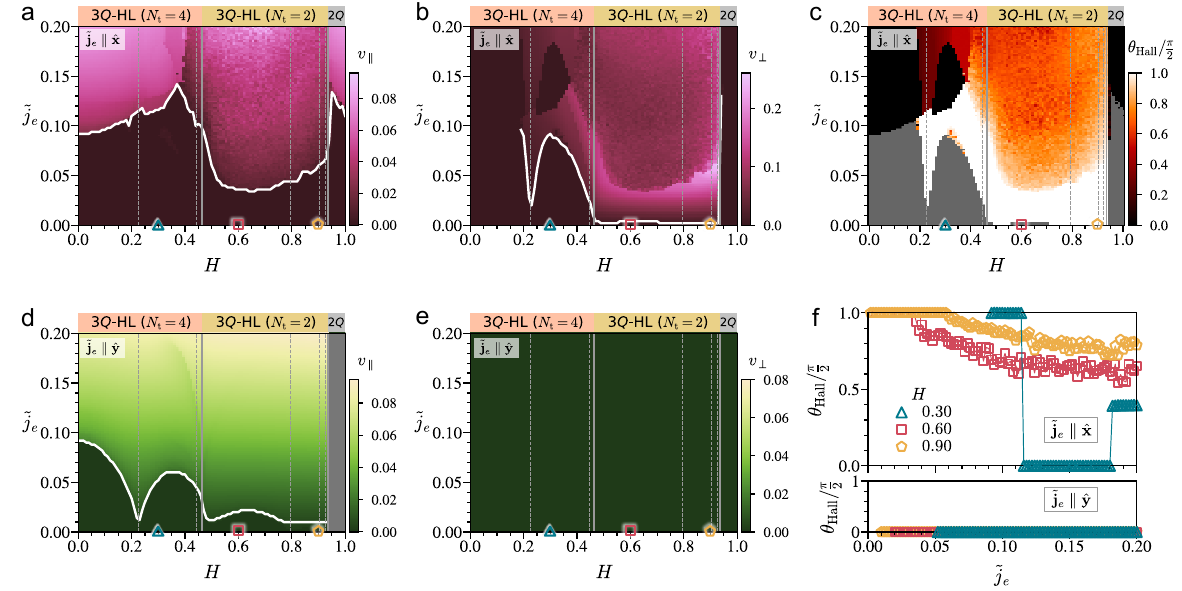}
\caption{
\label{fig:current_velocity}
\textbf{Toron Hall effect.} 
{\bf a,b} The velocities of spin textures for the longitudinal (\textbf{a}) and transverse motions (\textbf{b}) under the electron flow in the $x$ direction ($\tilde{\mathbf{j}}_e\parallel\hatv{x}$) on the plane of the magnetic field $H$ and the current density $\tilde{j}_e$. 
The white lines represent the threshold current densities, $\tilde{j}_{e,\mathrm{c}\parallel}$ (\textbf{a}) and $\tilde{j}_{e,\mathrm{c}\perp}$ (\textbf{b}). 
The dashed and solid gray vertical lines denote the values of $H$ where $d_{\mathrm{t}}$ and $N_{\mathrm{t}}$ show discontinuous changes, respectively; see Fig.~\figref{gs}{}. 
{\bf c} The Hall angle of the drift motion, $\thetaH$ [Eq.~\eqref{eq:Hall}]. 
In the black and white regions, the torons exhibit the zero and perfect Hall effects, respectively, while they are immobile in the gray region. 
{\bf d,e} The velocities of spin textures for the longitudinal (\textbf{d}) and transverse (\textbf{e}) motions under the electron flow in the $y$ direction ($\tilde{\mathbf{j}}_e\parallel\hatv{y}$). 
The latter is always zero in this parameter region, meaning that the zero toron Hall effect always occurs for $\tilde{j}_e > \tilde{j}_{e,\mathrm{c}\parallel}$ (above the white curve in \textbf{d}). 
$v_{\parallel}$ cannot be defined in the gray area. 
{\bf f} $\tilde{j}_e$ dependence of $\thetaH$ at three representative values of $H$ for $\tilde{\mathbf{j}}_e\parallel\hat{\mathbf{x}}$ (top) and $\tilde{\mathbf{j}}_e\parallel\hat{\mathbf{y}}$ (bottom). 
For $\tilde{\mathbf{j}}_e\parallel\hat{\mathbf{x}}$, the perfect toron Hall effect with $\theta_{\mathrm{Hall}} = \pi/2$ occurs for small $\tilde{\mathbf{j}}_e$, and it can be switched to the zero toron Hall effect with $\thetaH=0$ by increasing $\tilde{j}_e$ at $H=0.3$. 
In contrast, for $\tilde{\mathbf{j}}_e\parallel\hat{\mathbf{y}}$, only the zero toron Hall effect with $\thetaH = 0$ is observed. 
}
\end{figure*}

\begin{figure*}[tb]
\centering
\includegraphics[width=2.0\columnwidth]{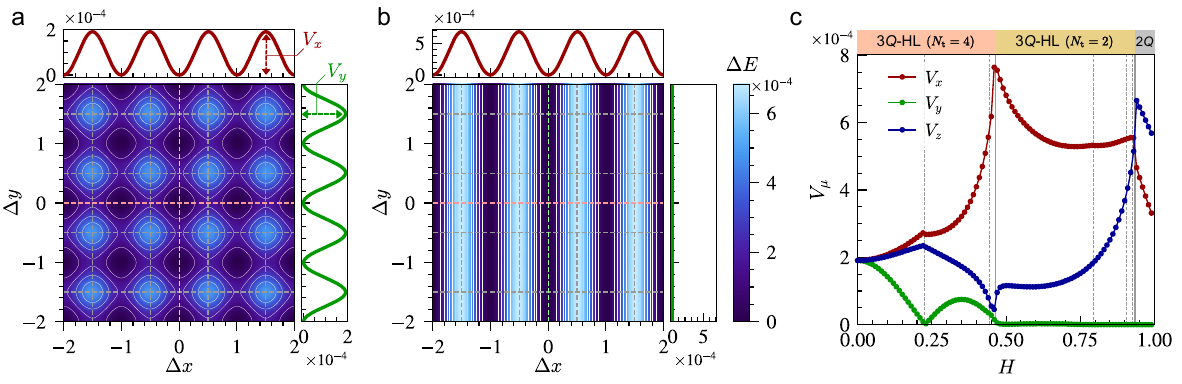}
\caption{
\label{fig:potential_barrier}
\textbf{Potential barrier for magnetic torons.} 
{\bf a,b} Energy change per site, $\Delta E$, for an in-plane spatial translation of the 3$Q$-HL by $\boldsymbol{\Delta}\mathbf{r} =(\Delta x, \Delta y, 0)$ measured from that for the ground state at $H=0$ (\textbf{a}) and $H=0.5$ (\textbf{b}). 
The potential is equivalent in the $x$ and $y$ directions at $H=0$, while it becomes almost one-dimensional at $H=0.5$. 
The gray dashed lines denote the intersections of the $xy$ plane and the unit cubes composed of the lattice sites. 
$\Delta E$ along the red and green dashed lines as well as the definitions of the potential barriers $V_x$ and $V_y$ are presented in the top
and right panels, respectively. 
{\bf c} $H$ dependence of the potential barriers in the direction of $\mu$, $V_{\mu}$. 
The behaviors of $V_x$ and $V_y$ qualitatively explain those of the threshold current densities of $v_{\parallel}$ for $\tilde{\mathbf{j}}_e \parallel \hatv{x}$ and $\hatv{y}$ in Figs.~\figref{current_velocity}{a} and \figref{current_velocity}{d}, respectively. 
}
\end{figure*}

To elucidate the current-induced dynamics of magnetic torons, we consider a spin model for a metallic chiral magnet on a simple cubic lattice, and perform a real-time simulation in an electric current based on the Landau-Lifshitz-Gilbert equation. 
The model includes effective interactions of Ruderman-Kittel-Kasuya-Yosida and Dzyaloshinskii-Moriya (DM) types, which has been shown to approximately reproduce the HLs discovered in MnSi$_{1-x}$Ge$_x$~\cite{Okumura2020,Kato2021, Kato2022} (see Methods). 
In the absence of the electric current, the model stabilizes a HL represented by the superposition of three spin density waves, called the $3Q$-HL, at zero magnetic field (Fig.~\figref{gs}{a}) (see Supplementary Note 1). 
The magnetic unit cell (MUC) contains eight Bloch points (hedgehogs and antihedgehogs), which are connected in four pairs by the skyrmion strings (Fig.~\figref{gs}{d}); namely, there are four torons denoted as $N_{\mathrm{t}} = 4$. 
In the magnetic field applied in the $z$ direction, $H$, the hedgehogs and antihedgehogs move toward each other along the skyrmion strings (Figs.~\figref{gs}{e},\textbf{f}), and the half of them disappear with pair annihilation at $H=H_{\mathrm{topo}}\simeq0.465$, leading to the topological transition to the $3Q$-HL with $N_{\mathrm{t}}=2$ (Figs.~\figref{gs}{b},\textbf{g}). 
By further increasing $H$, the system turns into a $2Q$ state at $H=H_{\mathrm{c}}\simeq0.935$, which is topologically trivial with no Bloch points ($N_{\rm t}=0)$ (Fig.~\figref{gs}{c}).

Let us first discuss drift motions caused by an electron flow $\tilde{\mathbf{j}}_e$ in the $x$ direction perpendicular to the skyrmion strings (see Methods). 
We find that the magnetic torons exhibit a transverse motion to the electron flow due to the Magnus force as well as a longitudinal one over the wide range of the HL phases under the magnetic field. 
We call this the toron Hall effect. 
Remarkably, however, the velocities of the longitudinal and transverse motions, $v_\parallel$ and $v_\perp$, respectively, show largely different and complicated dependences on $H$ and $\tilde{j}_e=|\tilde{\mathbf{j}}_e|$. 
Figures~\figref{current_velocity}{a} and \figref{current_velocity}{b} summarize $v_\parallel$ and $v_\perp$, respectively, on the $H$-$\tilde{j}_e$ 
plane. 
The longitudinal motion occurs above a certain threshold current density $\tilde{j}_{e,\mathrm{c}\parallel}$ represented by the white curve; $\tilde{j}_{e,\mathrm{c}\parallel}$ overall increases with $H$, except significant reduction near the topological transition at $H_{\mathrm{topo}}\simeq0.465$. 
Above $\tilde{j}_{e,\mathrm{c}\parallel}$, $v_\parallel$ monotonically increases with $\tilde{j}_e$. 
Meanwhile, the transverse motion also occurs above a threshold $\tilde{j}_{e,\mathrm{c}\perp}$, but it shows a sharp dip at $H\simeq 0.225$ and becomes vanishingly small for $H \gtrsim H_{\mathrm{topo}}$. 
The dip corresponds to a shortening of the skyrmion strings (Fig.~\figref{gs}); we will discuss this later. 
In addition, $v_\perp$ exhibits nonmonotonic behavior while increasing $\tilde{j}_e$ across $\tilde{j}_{e,\mathrm{c}\parallel}$; most significantly, around $H=0.3$, it becomes nonzero above $\tilde{j}_{e,\mathrm{c}\perp}$, but vanishes for $\tilde{j}_e > \tilde{j}_{e,\mathrm{c}\parallel} \sim 0.12$. 
We also note that $v_\perp$ vanishes in the weak field region where $H\lesssim 0.2$ in this range of $\tilde{j}_e$ due to the weak Magnus force stemming from small EMF (see Supplementary Note 1).

These distinct behaviors of $v_\parallel$ and $v_\perp$ indicate that the system exhibits the toron Hall effect in a wide range including two extreme limits. 
One is a purely transverse motion without any longitudinal one ($v_\parallel = 0$ and $v_\perp \neq 0$), appearing in the region where $\tilde{j}_{e,\mathrm{c}\perp} < \tilde{j}_e < \tilde{j}_{e,\mathrm{c}\parallel}$. 
We call this the {\it perfect} toron Hall effect. 
The other is opposite: a purely longitudinal motion accompanied with no transverse one ($v_\parallel \neq 0$ and $v_\perp=0$), occurring in the region where $v_\perp$ vanishes despite $\tilde{j}_e > \tilde{j}_{e,\mathrm{c}\parallel}$. 
We call this the {\it zero} toron Hall effect. 
These two extreme behaviors can be explicitly visualized by plotting the Hall angle defined by
\begin{align}
\thetaH = \arctan\left(\frac{v_{\perp}}{v_{\parallel}}\right), 
\label{eq:Hall}
\end{align}
which represents the net direction of the drift motion with respect to the electron flow (Fig.~\figref{current_velocity}{c}). 
The perfect and zero toron Hall effects correspond to $\thetaH=\pi/2$ and $\thetaH=0$, respectively, shown by the white and black regions. 
The former occurs in the wide range of $H \gtrsim 0.2$, especially for small $\tilde{j}_e$ in the $3Q$-HL with $N_{\mathrm{t}}=2$ after the topological transition, while the latter appears in the $3Q$-HL with $N_{\mathrm{t}}=4$ for $\tilde{j}_e > \tilde{j}_{e,\mathrm{c}\parallel}$.

Notably, one can achieve the zero toron Hall effect over a wider range of $H$ and $\tilde{j}_e$ simply by changing the current direction. 
We demonstrate this by taking the electron flow in the $y$ direction along the skyrmion strings ($\tilde{\mathbf{j}}_e \parallel \hatv{y}$). 
In this case also, the longitudinal motion occurs above nonzero threshold $\tilde{j}_{e,\mathrm{c}\parallel}$ (Fig.~\figref{current_velocity}{d}); $\tilde{j}_{e,\mathrm{c}\parallel}$ shows a different $H$ dependence from the $\tilde{\mathbf{j}}_e \parallel \hatv{x}$ case with characteristic dip structures, which will be discussed later. 
We note that $v_{\parallel}$ cannot be defined in the $2Q$ phase since the spin modulation in the $y$ direction is absent (see Supplementary Note 1). 
In stark contrast, the transverse motion is not observed at all in this parameter region (Fig.~\figref{current_velocity}{e}). 
Thus, with this current setting, the zero toron Hall effect appears in the entire region for $\tilde{j}_e > \tilde{j}_{e,\mathrm{c}\parallel}$. 

Our results show that the toron Hall effect is much more flexible than the skyrmion Hall effect. 
It can be controlled by the amplitudes and directions of the magnetic field and the electric current. 
Most strikingly, it includes two extremes, the zero and perfect toron Hall effects, which have never been reported in the skyrmion Hall effect. 
By choosing appropriate parameters, one can even achieve switching between the two extremes, as demonstrated in Fig.~\figref{current_velocity}{f}: 
By increasing $\tilde{j}_e$ for $\tilde{\mathbf{j}}_e\parallel \hatv{x}$ at $H=0.3$, the perfect toron Hall effect is observed for $0.091 \lesssim \tilde{j}_e \lesssim 0.115$ (Supplementary Video 1), followed by the zero toron Hall effect for $0.115 \lesssim \tilde{j}_e \lesssim 0.181$ (Supplementary Video 2), and finally, the toron Hall effect with $\theta_{\mathrm{Hall}} \sim \frac{\pi}{5}$ appears for $\tilde{j}_e \gtrsim 0.181$ (Supplementary Video 3).

We note that the toron Hall effect vanishes when we turn off the DM-type interaction in our model for the $3Q$-HL (see Supplementary Note 5). 
Nonetheless, this does not mean that the toron Hall effect never occur in nonchiral magnets. 
In general, the net EMF can be nonzero even in nonchiral cases in the presence of spin-orbit coupling, leading to the topological Hall effect as observed in SrFeO$_3$~\cite{Ishiwata2011}, allowing the toron Hall effect as its counteraction.

\section*{Anisotropic potential barrier \label{sec:potential}}

The peculiar toron Hall effect originates from the energy potential for torons on the discrete lattice. 
Drift motions of nanoscale torons are strongly affected by the lattice discretization due to their short magnetic pitch and singular structures of constituting Bloch points. 
The magnetic field influences the spin textures, resulting in energy potential modulation in an anisotropic manner. 
This causes the different and complicated $H$ dependences of $\tilde{j}_{e,\mathrm{c}\parallel}$ and $\tilde{j}_{e,\mathrm{c}\perp}$, as demonstrated below.

When $H=0$, the energy potential for torons on the $xy$ plane is fourfold rotational symmetric reflecting the lattice symmetry (Fig.~\figref{potential_barrier}{a}). 
Here, we plot the energy increase per site $\Delta E$ by shifting the spin texture with $\boldsymbol{\Delta} \mathbf{r}=(\Delta x, \Delta y, 0)$ from the ground state. 
The potential oscillates with the lattice period, and the heights of the potential barrier in the $x$ and $y$ directions are equivalent, $V_x=V_y$. 
In this case, an electric current required for torons to overcome the potential barrier is equal for $\tilde{\mathbf{j}}_e \parallel \hatv{x}$ and $\tilde{\mathbf{j}}_e \parallel \hatv{y}$, resulting in $\tilde{j}_{e,\mathrm{c}\parallel} \simeq 0.09$ in both cases (Figs.~\figref{current_velocity}{a},\textbf{d}). 
By introducing $H$, $V_x$ increases but $V_y$ decreases; namely, the potential develops a one-dimensional anisotropy through the modulation of the spin texture, giving rise to different threshold current densities for the different current directions.  
Figure~\figref{potential_barrier}{b} represents an example at $H=0.5$, where $V_y$ almost vanishes. 
Such a highly anisotropic potential causes the unique drift motions including the zero and perfect toron Hall effects found above. 
On one hand, for $\tilde{\mathbf{j}}_e \parallel \hatv{x}$, the longitudinal drift motion is hindered by large $V_x$, while the transverse motion is easily driven with vanishingly small $V_y$, resulting in $\tilde{j}_{e,\mathrm{c}\perp} < \tilde{j}_{e,\mathrm{c}\parallel}$. 
This leads to the perfect toron Hall effect, as illustrated in Fig.~\figref{schematic}{d}. 
On the other hand, for $\tilde{\mathbf{j}}_e \parallel \hatv{y}$, the longitudinal drift motion is caused by a small current density, while the transverse motion is largely suppressed, resulting in $\tilde{j}_{e,\mathrm{c}\parallel} < \tilde{j}_{e,\mathrm{c}\perp}$. 
This brings about the zero toron Hall effect, as illustrated in the left panel of Fig.~\figref{schematic}{e}. 
The zero toron Hall effect for $\tilde{\mathbf{j}}_{e} \parallel \hatv{x}$ in Fig.~\figref{current_velocity}{c} is also caused by the anisotropic potential, but in more complicated competition between the longitudinal driving force from the electric current and the induced transverse Magnus force (the right panel of Fig.~\figref{schematic}{e}).

Remarkably, the energy potential becomes almost one-dimensional ($V_y \simeq 0$) in the entire $3Q$-HL phase with $N_{\mathrm{t}} = 2$ after the topological transition (Fig.~\figref{potential_barrier}{c}). 
The origin of the vanishingly small $V_y$ lies in the suppression of the higher harmonics of the spin structure factor in the $y$ direction; the suppression beyond the Nyquist wave number makes the spin texture less susceptible to the lattice discretization (see Supplementary Note 2). 
It is worth noting that the $H$ dependences of the potential barriers qualitatively explain those of the critical current densities: $V_x$ behaves as $\tilde{j}_{e,\mathrm{c}\parallel}$ for $\tilde{\mathbf{j}}_e \parallel \hatv{x}$ (Fig.~\figref{current_velocity}{a}), and $V_y$ as $\tilde{j}_{e,\mathrm{c}\perp}$ for $\tilde{\mathbf{j}}_e \parallel \hatv{x}$ (Fig.~\figref{current_velocity}{b}) and $\tilde{j}_{e,\mathrm{c}\parallel}$ for $\tilde{\mathbf{j}}_e \parallel \hatv{y}$ (Fig.~\figref{current_velocity}{d}) (for $V_z$, see Supplementary Note 3). 
This supports our argument based on the potential barrier on the discrete lattice. 
We note that $V_x$ does not follow the significant reduction of $\tilde{j}_{e,\mathrm{c}\parallel}$ for $\tilde{\mathbf{j}}_e \parallel \hatv{x}$ near the topological transition in Fig.~\figref{current_velocity}{a}, but this is accounted for by the Magnus force in the direction longitudinal to the current arising from the transverse drift motion of torons.

\section*{Probing hidden topology \label{sec:probing}}

Our results indicate that the $H$ dependences of the threshold current densities, $\tilde{j}_{e,\mathrm{c}\parallel}$ and $\tilde{j}_{e,\mathrm{c}\perp}$, can be a sensitive probe of the topological transition with toron 
annihilation. 
This can be complementary to the elastic constant measurements~\cite{Kanazawa2016}. 
Furthermore, they are useful for probing more detailed changes in the topological objects, namely the length changes in the skyrmion strings. 
For instance, for $\tilde{\mathbf{j}}_e \parallel \hatv{x}$, $\tilde{j}_{e,\mathrm{c}\perp}$ exhibits a sharp dip at $H\simeq 0.225$ (Fig.~\figref{current_velocity}{b}), where the lengths of the skyrmion strings change from $d_{\mathrm{t}} = 7$ to $6$ in two out of four torons and from $5$ to $4$ in the rest (Fig.~\figref{gs}). 
We also observe a small anomaly in $\tilde{j}_{e,\mathrm{c}\parallel}$ (Fig.~\figref{current_velocity}{a}). 
A similar sharp dip is also found in $\tilde{j}_{e,\mathrm{c}\parallel}$ for $\tilde{\mathbf{j}}_e \parallel \hatv{y}$ (Fig.~\figref{current_velocity}{d}), but in this case, a smaller dip is additionally seen at $H\simeq 0.765$ in the $3Q$-HL with $N_{\mathrm{t}}=2$, where the length of the skyrmion string reduces from $6$ to $5$ in the remaining two torons.
These behaviors are understood as follows. 
The Bloch points locate at the interstitial positions on the lattice so that their cores with vanishing spin length avoid the lattice sites, and hence the energy potential is strongly modulated when the Bloch points traverse between the unit cubes of the lattice with the shortening of the skyrmion strings. 
Thus, the measurement of the threshold current densities can serve as a probe of not only topological transitions but also detailed changes of topological objects in the lattice spacing scale that are usually hidden in the macroscopic properties~\cite{Kato2022,Kato2023}. 
We note that this holds even for nonchiral cases without the DM-type interaction (see Supplementary Note 5). 
 
In experiments, the threshold current densities under the magnetic field can be measured straightforwardly. 
In the present simulation, $\tilde{j}_e \sim 0.1$ corresponds to the electric current density of $\sim 10^{12}$~A/m$^{2}$, when assuming the energy unit as $1$~meV, the lattice constant as $0.5$~nm, and the spin polarization of the current as $0.2$~\cite{Iwasaki2013} (see Methods). 
This is comparable to a typical value required for magnetic domain walls to drive~\cite{Parkin2008}. 
Thus, the threshold current measurements would provide microscopic information of the topology, complementary to the real-space measurements such as the Lorentz transmission electron microscopy~\cite{Tanigaki2015} and 3D tomographic imaging techniques~\cite{Donnelly2017,Hierro2020}.
Such measurements for nanoscale HLs discovered in MnSi$_{1-x}$Ge$_{x}$ and SrFeO$_3$ are eagerly awaited for the verification of our theoretical prediction.

\section*{Concluding remarks}

Our comprehensive study of the current-induced dynamics of nanoscale magnetic torons has revealed the extensive controllability of their drift motions spanning from the zero to perfect toron Hall effect. 
We clarified that such prominent current-induced dynamics arises from the potential barrier on the discrete lattice which can be highly anisotropic through the modulation of the spin textures by the magnetic field. 
We also found that the field dependence of the threshold current serves as a concise and sensitive probe of the microscopic changes of topological objects.

While the measurement of the threshold currents is experimentally straightforward, their actual values as well as the $H$ dependences depend on the crystal structure and its symmetry of the real materials. 
Our results for the simple cubic lattice need modifications to apply to MnGe with the $B20$ structure, while they will be rather straightforwardly extended to the cubic perovskite SrFeO$_3$.
Nevertheless, we believe that the flexible and controllable toron Hall effect in the wide range of the Hall angle is universally observed for HLs with nanoscale torons because of the short magnetic pitch comparable to the lattice constant and the singularity at the cores of the Bloch points. 
Extensions of our studies to crystallographic groups other than cubic are intriguing for further explorations. 
For instance, hexagonal systems may allow zero and perfect toron Hall effects even at zero field because of the symmetry of the energy potential.

Magnetic torons are promising information carriers due to their topological robustness. 
Our finding of flexible and controllable drift motions will increase their advantage over skyrmions. 
In the case of skyrmions, it was shown that the current-driven motions are almost free from the lattice discretization, and the Hall motion inevitably occurs, which has hindered the applications of skyrmions to spintronic devices~\cite{Jonietz2010,Zang2011,Iwasaki2013}. 
The extremely high controllability of the toron Hall effect is expected to resolve these issues and lead to future device applications. 
We note, however, that such a skyrmion motion is partly due to the real-space scale of spin textures much larger than the lattice spacing. 
Recently, nanoscale skyrmions have been discovered beyond the conventional DM mechanism~\cite{Kurumaji2019,Hirschberger2019,Khanh2020,Takagi2022}. 
Such short-pitch skyrmions would be more susceptible to the lattice discretization compared to conventional long-pitch ones, while they are anticipated to be less sensitive than magnetic torons due to the absence of singularities. 
Detailed comparisons, including other topological spin textures, will be left to future studies.

\begin{acknowledgments}
The authors thank R. Arita, G.-W. Chern, N. Kanazawa, K. Kobayashi, M. Mochizuki, R. Takagi, and H. Yoshimochi for fruitful discussions.
This research was supported by Grant-in-Aid for Scientific Research Grants (Nos. JP19H05822, JP19H05825, JP21J20812, JP22K03509, JP22K13998, and JP23H01119), JST CREST (Nos. JP-MJCR18T2 and JP-MJCR19T3), and the Chirality Research Center in Hiroshima University and JSPS Core-to-Core Program, Advanced Research Networks. K.S. was supported by the Program for Leading Graduate Schools (MERIT-WINGS). Parts of the numerical calculations were performed in the supercomputing systems in ISSP, the University of Tokyo.
\end{acknowledgments}

\appendix

\section*{Methods} 
\label{sec:method}

\subsection*{Model for metallic chiral magnets}
\label{sec:model}

We adopt an effective spin model for metallic chiral magnets with itinerant electrons coupled to localized spins. 
The electron-spin coupling brings about effective long-range interactions between the localized spins known as the Ruderman-Kittel-Kasuya-Yosida interaction~\cite{Ruderman1954, Kasuya1956, Yosida1957}. 
By extending the argument, the previous studies systematically derived the generalized and multiple-spin interactions in the presence of the spin-orbit coupling, and showed that such extensions stabilize HLs as well as skyrmion lattices with short magnetic pitches~\cite{Hayami2017,Hayami2018,Okada2018,Okumura2020,Kato2021,Okumura2022,Kato2022,Kato2023}. 
In this study, we consider the symmetry-adapted two-spin interactions for a cubic chiral magnet by following the previous studies~\cite{Kato2021,Kato2022}. 
The Hamiltonian reads 
\begin{eqnarray}
\mathcal{H}=\sum_{\substack{i,j \\ g_3({\bf r}_{ij})\leq r_{\rm c}}}
{\bf S}_{i}^{{\mathsf T}}(t){\mathsf J}_{ij}{\bf S}_{j}(t)
-\sum_{i}H S_{i,z}(t), 
\label{eq:Heff_bil_3D}
\end{eqnarray}
where 
\begin{align}
&g_{3}({\bf r})=|x|+|y|+|z|, \\
&\begin{alignedat}{2}
{\mathsf J}_{ij}=\frac{2e^{-\gamma g_3({\bf r}_{ij})}}{\{L_{\gamma}(0)\}^3}\sum_{\eta=1}^{3}
&\left[{\rm Re}{\mathsf B}_{{\bf Q}_{\eta}} \left(
\cos({\bf Q}_{\eta}\cdot{\bf r}_{ij})
- \bar{L}_{\gamma}({\bf Q}_{\eta})
\right) \right. \\
& \left. -{\rm Im}{\mathsf B}_{{\bf Q}_{\eta}} \sin({\bf Q}_{\eta}\cdot{\bf r}_{ij})\right], 
\end{alignedat} 
\end{align}
with
\begin{align}
&\mathsf{B}_{{\bf Q}_1}=
\left(\begin{array}{ccc}
-J(1-\Delta) & 0 & 0 \\
0 & -J(1+2\Delta) & \mathrm{i}D \\
0 & -\mathrm{i}D & -J(1-\Delta)
\end{array}\right), \\
&\mathsf{B}_{{\bf Q}_2}=
\left(\begin{array}{ccc}
-J(1-\Delta) & 0 & -\mathrm{i}D \\
0 & -J(1-\Delta) & 0 \\
\mathrm{i}D & 0 & -J(1+2\Delta)
\end{array}\right), \\
&\mathsf{B}_{{\bf Q}_3}=
\left(\begin{array}{ccc}
-J(1+2\Delta) & \mathrm{i}D & 0 \\
-\mathrm{i}D & -J(1-\Delta) & 0 \\
0 & 0 & -J(1-\Delta)
\end{array}\right),\\
&\bar{L}_{\gamma}({\bf q})=
\frac{L_{\gamma}(q_x)}{L_{\gamma}(0)}
\frac{L_{\gamma}(q_y)}{L_{\gamma}(0)}
\frac{L_{\gamma}(q_z)}{L_{\gamma}(0)}, \\
&L_{\gamma}(q)=\frac{\sinh\gamma}{\cosh\gamma - \cos q}. 
\end{align}
Here, $\mathbf{S}_i(t)$ denotes the classical spin at the site $i$ located at $\mathbf{r}_i$ and time $t$ ($|\mathbf{S}_i(t)|=1$); $J$, $\Delta$, and $D$ represent the coefficients for the isotropic exchange, anisotropic exchange, and DM interactions, respectively. 
The interactions exhibit spatial oscillations arising from the itinerant nature of electrons; we set the wave vectors $\mathbf{Q}_1=(Q,0,0)$, $\mathbf{Q}_2=(0,Q,0)$, and $\mathbf{Q}_3=(0,0,Q)$, which are related with each other by threefold rotational symmetry about the [111] axis.  
Furthermore, we introduce the spatial decay of the interaction parametrized by $\gamma$ as a function of $\mathbf{r}_{ij}=\mathbf{r}_j-\mathbf{r}_i$ and reduce the numerical cost by limiting the summation in real space for $g_3(\mathbf{r}_{ij})\leq r_{\mathrm{c}}$ by following the previous studies~\cite{Kato2021,Shimizu2023crystallization}. 
The second term in Eq.~\eqref{eq:Heff_bil_3D} represents the Zeeman coupling to the magnetic field $\mathbf{H}=(0,0,H)$. 
We take $J$ as the energy unit, set the lattice constant unity, and choose $\Delta=0.3$, $D=0.3$, $Q=\pi/6$, $\gamma=0.6Q$, and $r_{\mathrm{c}}=16$. 
We perform the calculations for the $N=12^3$-site system under the periodic boundary condition, which corresponds to the single MUC. 
We confirm that the results are qualitatively intact for $N=60^3$. 

\subsection*{Landau-Lifshitz-Gilbert equation}
\label{sec:LLG}

To elucidate the current-induced dynamics of the HL, we study the real-time dynamics of spins by solving the Landau-Lifshitz-Gilbert (LLG) equation given by
\begin{align}
\frac{d \mathbf{S}_i(t)}{dt} = -\mathbf{S}_i(t)\times\mathbf{h}_i^{\mathrm{eff}}(t)-\alpha\mathbf{S}_i(t)\times\frac{d \mathbf{S}_i(t)}{dt}+\boldsymbol{\tau}_i(t),
\label{eq:LLGj}
\end{align}
where $\alpha$ is the Gilbert damping, ${\bf h}^{\rm eff}_i(t)$ is the mean magnetic field at time $t$, defined as 
\begin{eqnarray}
{\bf h}^{\rm eff}_i(t) = \frac{\partial \mathcal{H}(t)}{\partial {\bf S}_i(t)}. 
\label{eq:heff}
\end{eqnarray}
Note that the length constraint of $|{\bf S}_i(t)|=1$ is deferred only in the calculation of this derivative. 
In Eq.~\eqref{eq:LLGj}, $\boldsymbol{\tau}_i(t)$ is the torque arising from an electric current, which is given by~\cite{Zhang2004,Iwasaki2013} 
\begin{align}
\boldsymbol{\tau}_i(t)=-\left[
\left(\tilde{\mathbf{j}}_e\cdot\hat{\boldsymbol{\delta}}\right)\mathbf{S}_i(t)
+\beta\left(\mathbf{S}_i(t)\times\left(\tilde{\mathbf{j}}_e\cdot\hat{\boldsymbol{\delta}}\right)\mathbf{S}_i(t)\right)
\right].
\label{eq:tau}
\end{align}
Here, the effect of the electric current is incorporated as the spin transfer torque; $\tilde{\mathbf{j}}_e$ represents the normalized velocity of conduction electrons,
\begin{align}
\tilde{\mathbf{j}}_e=-\frac{pa^2}{2e}\mathbf{j},
\end{align}
where $p$ is the spin polarization, $a$ is the lattice constant (we set $a=1$), $e>0$ is the elementary charge, and $\mathbf{j}$ is the electric current. 
In Eq.~\eqref{eq:tau}, $\beta$ is the nonadiabatic coefficient, and $\hat{\boldsymbol \delta}=(\hat{\delta}_x, \hat{\delta}_y, \hat{\delta}_z)$ is introduced to approximate the spatial derivative on the discrete lattice as 
\begin{eqnarray}
\hat{\delta}_{\mu}{\bf S}_{{\bf r}_i}(t)=\frac{1}{2}\left({\bf S}_{{\bf r}_i+\hat{\boldsymbol \mu}}-{\bf S}_{{\bf r}_i-\hat{\boldsymbol \mu}}\right).  
\end{eqnarray}

In this study, we numerically solve Eq.~\eqref{eq:LLGj} by using the fourth-order Runge-Kutta method with the time step $\Delta t=0.1$. 
We take $\alpha=0.04$ and $\beta=0.02$, which are typical values for ferromagnets~\cite{
Zhang2004,Oogane2006}.

\subsection*{Calculating velocity}
\label{sec:velocity}

We calculate the velocities of drift motions of the spin textures by using the time series of the Fourier components of spins. 
By assuming that spin structures in continuum space exhibit the rigid drift motion, a real-space and real-time spin configuration can be given by $\mathbf{S}(\mathbf{r}, t+\Delta t)=\mathbf{S}(\mathbf{r}-\mathbf{v}\Delta t, t)$, where $\mathbf{S}(\mathbf{r}, t)$ is the spin at the position $\mathbf{r}$ and time $t$, and $\mathbf{v}$ is the velocity of the spin texture. 
The Fourier transform of spins is then given by 
\begin{align}
\mathbf{S}(\mathbf{q},t+\Delta t)=\int\mathbf{S}(\mathbf{r},t+\Delta t)e^{-\mathrm{i}\mathbf{q}\cdot\mathbf{r}}d^3\mathbf{r}=\mathbf{S}(\mathbf{q},t)e^{-\mathrm{i}\mathbf{q}\cdot\mathbf{v}\Delta t}, 
\end{align}
where $\mathbf{q}$ is the wave vector. 
Hence, the velocity at $t$ can be calculated by the phase difference between $\mathbf{S}(\mathbf{q},t)$ and $\mathbf{S}(\mathbf{q},t+\Delta t)$. 
In the actual calculations, we use the above relation for the Fourier components for the ordering wave vectors $\mathbf{Q}_1$, $\mathbf{Q}_2$, and $\mathbf{Q}_3$ calculated from $\mathbf{S}_i(t)$ obtained by the LLG simulation as 
\begin{align}
v_\mu=-\frac{1}{Q(t_f-t_i)}\int_{t_i}^{t_f}~\frac{d}{dt}\arg\left(S_{{\bf Q}_\eta, \mu^\prime}(t)\right)dt, 
\label{eq:v}
\end{align}
where $\eta=1,2,3$ for $\mu=x,y,z$, respectively, and $\mu^\prime$ denotes the largest component of ${\bf S}_{{\bf Q}_{\eta}}$ determined by the exchange anisotropy in the model in Eq.~\eqref{eq:Heff_bil_3D}. 
For instance, at ${\bf q}={\bf Q}_1$, the exchange anisotropy is strong in the $S_y$ direction, leading to the largest amplitude in $S_{{\bf Q}_1,y}$. 
The integral in Eq.~\eqref{eq:v} is taken from $t_i=500$ to $t_f=1000$.

\subsection*{Calculating potential barrier}
\label{sec:barrier_calculation}

We compute the potential barrier for magnetic torons by the energy change with shifting their positions from the ground state. 
For this calculation, we first approximately express the ground-state spin configuration by a superposition of three elliptical spirals using 
\begin{eqnarray}
{\bf S}_i^{\mathrm{ellip}3Q}\propto\left(
\begin{array}{c}
\psi_2\varepsilon_2\sin\QQ_2 + \psi_3\cos\QQ_3 \\
\psi_1\cos\QQ_1 + \psi_3\varepsilon_3\sin\QQ_3 \\
\psi_1\varepsilon_1\sin\QQ_1 + \psi_2\cos\QQ_2 + m
\end{array}
\right),
\label{eq:ellip3Q}
\end{eqnarray}
where $\QQ_{\eta}={\bf Q}_{\eta}\cdot{\bf r}_i+\vp_{\eta}$, $\psi_{\eta}$ and $\varepsilon_{\eta}$ respectively represent the long axis and the ellipticity of the constituting spirals, and $\vp_{\eta}$ is the phase.  
We estimate these parameters by minimizing the cost function defined by 
\begin{align}
&U(\{\psi_{\eta}, \varepsilon_{\eta}, \vp_\eta\}, m)=\notag \\
&\frac{1}{N}\sum_{i} 
\left(1-{\bf S}^{\rm LLG}_{i} \cdot {\bf S}^{{\rm ellip}3Q}_{i}(\{\psi_{\eta}, \varepsilon_{\eta}, \vp_\eta\}, m) \right), 
\label{eq:cost} 
\end{align}
where ${\bf S}_i^{{\rm LLG}}$ is the ground-state spin configuration obtained by numerically solving the LLG equation and ${\bf S}_i^{{\rm ellip}3Q}$ is that generated from Eq.~\eqref{eq:ellip3Q}. 
The optimal state $\{{\bf S}_i^{{\rm ellip}3Q}\}^*$ shows the cost function less than $6\times10^{-4}$ for all $H$, and hence $\{{\bf S}_i^{\rm LLG}\}$ is well reproduced by Eq.~\eqref{eq:ellip3Q} (see Supplementary Note 4). 

Then, once the parameters in Eq.~\eqref{eq:ellip3Q} is obtained by minimizing Eq.~\eqref{eq:cost}, we introduce a spatial translation by $\Delta {\bf r}$ through a phase shift from $\vp_{\eta}$ to $\vp_{\eta} - \Delta \vp_{\eta}$~\cite{Shimizu2022phase}. 
Specifically, for the present $3Q$-HL case, a spatial translation by $\Delta {\bf r}=(\Delta x, \Delta y, \Delta z)$ is represented by a phase shift as 
\begin{eqnarray}
\Delta {\bf r}=\frac{1}{Q}\left(\Delta \vp_1, \Delta \vp_2, \Delta \vp_3 \right). 
\label{eq:Deltar}
\end{eqnarray}
Thus, by shifting the phase $\vp_{\eta}$ in the optimal spin state $\{{\bf S}_i^{{\rm ellip}3Q}\}^*$, we compute the energy change per site for the model in Eq.~\eqref{eq:Heff_bil_3D} measured from that for the optimal spin configuration $\{{\bf S}_i^{{\rm ellip}3Q}\}^*$, denoted by $\Delta E$. 

In the ground state, the Bloch points are located at the interstitial positions of the lattice sites, and the energy increases when they traverse across the unit cubes composed of the lattice sites. 
This implies that the $3Q$-HL undergoes the potential energy approximately given by 
\begin{align}
\Delta E(\Delta{\bf r}) = \sum_{\mu=x,y,z} \frac{V_\mu}{2}(1 - \cos(2\pi \Delta \mu)). 
\label{eq:cos_potential}
\end{align}
By assuming this form, we obtain the potential barrier $V_{\mu}$ as $V_{\mu}=\Delta E(\hat{\boldsymbol{\mu}}/2)$. 

\bibliography{ref}

\end{document}


\title{
Supplementary Information for 
``Current-induced motion of nanoscale magnetic torons over the wide range of the Hall angle''
}

\author{Kotaro Shimizu}
\affiliation{RIKEN Center for Emergent Matter Science, Saitama 351-0198, Japan}

\author{Shun Okumura} 
\affiliation{Department of Applied Physics, The University of Tokyo, Tokyo 113-8656, Japan}

\author{Yasuyuki Kato}
\affiliation{Department of Applied Physics, University of Fukui, Fukui 910-8507, Japan}

\author{Yukitoshi Motome}
\affiliation{Department of Applied Physics, The University of Tokyo, Tokyo 113-8656, Japan}

\date{\today}


\maketitle

\section{Ground-state property \label{sec:gs}}

\begin{figure*}[tb]
\centering
\includegraphics[width=0.7\columnwidth]{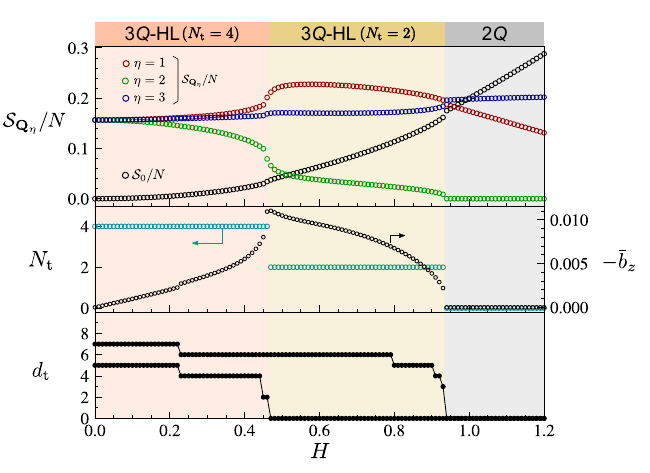}
\caption{
\label{fig:gs_hdep}
\textbf{Magnetic and topological properties in the ground state.} 
$H$ dependences of the spin structure factor $\mathcal{S}_{{\bf q}}$ (top), the number of torons within the MUC, $\Nt$, and the average EMF $\bar{b}_z$ (middle), and the length of the skyrmion string projected onto the $y$ axis, $\dt$ (bottom). 
}
\end{figure*}

We present the detailed properties of the ground states for the model used in the main text. 
The model exhibits spin textures that can be regarded as superpositions of spin density waves with the ordering vectors $\mathbf{Q}_\eta$. 
To identify them, we compute the spin structure factor $\SSFq$ defined by 
\begin{align}
&\SSFq = \frac{1}{N}\sum_{i,j} \mathbf{S}_i\cdot\mathbf{S}_j e^{\mathrm{i}\mathbf{q}\cdot(\mathbf{r}_i-\mathbf{r}_j)}.
\label{eq:SSF}
\end{align}
In addition, to characterize their topological property, we compute real-space positions of Bloch points marked by nonzero monopole charge $Q_{\mathrm{m}}$. 
On the cubic lattice in this study, $Q_{\mathrm{m}}$ is computed for each unit cube by~\cite{Okumura2020,Okumura2020tracing}
\begin{align}
Q_{\mathrm{m}}(\mathbf{r}_{\mathrm{c}})=
\frac{1}{4\pi}\sum_{\mu,\delta_{\mu}}\delta_{\mu}\Omega_{\square}\left(\mathbf{r}_{\mathrm{c}}+\frac{\delta_{\mu}}{2}\hatbsymbol{\mu}\right), 
\label{eq:Qm}
\end{align}
where $\delta_{\mu}=\pm1$, $\mathbf{r}_{\mathrm{c}}$ denotes the center of the unit cube, and $\hatbsymbol{\mu}$ is the unit vector in the $x$, $y$, or $z$ direction. 
Here, $\Omega_{\square}(\mathbf{r})$ represents the solid angle spanned by the four spins located at the vertices of the face centered at $\mathbf{r}$ among the six faces of the cube. 
To calculate $\Omega_{\square}(\mathbf{r})$, we divide each face into two triangles by the line parallel to $\hatbsymbol{\nu}+\hatbsymbol{\rho}$ with $\hatbsymbol{\mu}=\hatbsymbol{\nu}\times\hatbsymbol{\rho}$ ($\mu,\nu,\rho=x,y,z$), and take the sum of the solid angles for each triangle spanned by the three spins, which is given by 
\begin{align}
\Omega_{\triangle}=2\arctan\left(\frac{\mathbf{S}_a\cdot(\mathbf{S}_b\times\mathbf{S}_c)}{1+\mathbf{S}_a\cdot\mathbf{S}_b+\mathbf{S}_b\cdot\mathbf{S}_c+\mathbf{S}_c\cdot\mathbf{S}_a}\right). 
\end{align}
Here, $\mathbf{S}_a$, $\mathbf{S}_b$, and $\mathbf{S}_c$ are arranged in counterclockwise order when viewed from the direction of the $\mu$ axis, and the range of $\arctan$ is chosen to ensure $\Omega_{\triangle}\in [-2\pi, 2\pi]$. 
For the present model, $Q_{\mathrm{m}}(\mathbf{r}_{\mathrm{c}})$ in Eq.~\eqref{eq:Qm} takes $0$, $-1$, or $1$.  
We further compute the total number of torons within the magnetic unit cell (MUC), $N_{\mathrm{t}}$, by
\begin{align}
N_{\mathrm{t}} = \frac{1}{2} \sum_{\mathbf{r}_{\mathrm{c}}} |Q_{\mathrm{m}}(\mathbf{r}_{\mathrm{c}})|. 
\label{eq:Nt} 
\end{align}
We also calculate the average EMF $\bar{\mathbf{b}}$ by taking a spatial average of the EMF at site $i$, $\mathbf{b}_i=(b_{i,x}, b_{i,y}, b_{i,z})$, defined by
\begin{align}
b_{i,\mu} = \frac{1}{4}\sum_{\nu\,\rho,\delta_{\nu},\delta_{\rho}}
\epsilon_{\mu\nu\rho}\delta_{\nu}\delta_{\rho}
\mathbf{S}(\mathbf{r}_i)\cdot\left[
\mathbf{S}(\mathbf{r}_i+\delta_{\nu}\hatbsymbol{\nu})
\times
\mathbf{S}(\mathbf{r}_i+\delta_{\rho}\hatbsymbol{\rho})
\right], 
\label{eq:bi}
\end{align}
where $\mathbf{S}(\mathbf{r}_i)$ denotes the spin at position $\mathbf{r}_i$ [$\mathbf{S}(\mathbf{r}_i)=\mathbf{S}_i$] and $\epsilon_{\mu\nu\rho}$ is the Levi-Civita symbol. 
In addition, we also identify the skyrmion strings connecting the hedgehog and antihedgehog, by evaluating the vorticity $\zeta$ for six faces composing each unit cube. 
The vorticity for the face of the $i$-th unit cube on the $\mu\nu$ plane is given by
\begin{align}
\zeta=\frac{1}{2\pi}\left(
\argS{\mathbf{r}_i}{\mathbf{r}_i+\hatbsymbol{\mu}}
+\argS{\mathbf{r}_i+\hatbsymbol{\mu}}{\mathbf{r}_i+\hatbsymbol{\mu}+\hatbsymbol{\nu}}
+\argS{\mathbf{r}_i+\hatbsymbol{\mu}+\hatbsymbol{\nu}}{\mathbf{r}_i+\hatbsymbol{\nu}}
+\argS{\mathbf{r}_i+\hatbsymbol{\nu}}{\mathbf{r}_i}
\right), 
\label{eq:vorticity}
\end{align}
with $S^{+}(\mathbf{r}_i)=S_x(\mathbf{r}_i) + \mathrm{i}S_y(\mathbf{r}_i)$. 
The skyrmion string is supposed to run perpendicular to a face when $\zeta$ is nonzero and the average of $S_{i,z}$ for the four spins constituting the face is negative. 
Then, we compute the length of the skyrmion string projected onto the $y$ axis, $\dt$, for each toron.

The field dependences of the above quantities are presented in Fig.~\figref{gs_hdep}{}. 
The spin structure factor $\SSF{\mathbf{Q}_{\eta}}$ becomes nonzero for all three components  for $0 \leq H \lesssim 0.935$ and for two out of them for $H \gtrsim 0.935$; hence, we identify the former and latter as the $3Q$ and $2Q$ states, respectively. 
In the $3Q$ state, three $\SSF{\mathbf{Q}_{\eta}}$ are identical at zero field, but become inequivalent as $\SSF{\mathbf{Q}_{1}} > \SSF{\mathbf{Q}_{3}} > \SSF{\mathbf{Q}_{2}} > 0$ for nonzero fields. 
While increasing $H$, $\SSF{{\bf Q}_{\eta}}$ as well as the uniform component $\SSF{0}=\SSF{\mathbf{q}=0}$ exhibits an anomaly at $H = H_{\mathrm{topo}} \simeq 0.465$. 
At $H = H_{\mathrm{topo}}$, $\Nt$ changes from 4 to 2 through the annihilation of half of the torons, as plotted in the middle panel of Fig.~\ref{fig:gs_hdep}. 
Hence, the anomaly observed in $\SSFq$ signals this topological transition from the $3Q$-hedgehog lattice (HL) with $\Nt=4$ to the $3Q$-HL with $\Nt=2$. 
With further increasing $H$, $\Nt$ vanishes at $H\simeq 0.935$ where the $3Q$-HL with $\Nt=2$ turns into the trivial $2Q$ state. 

Regarding the EMF, $\bar{\mathbf{b}}=0$ at zero field due to the symmetry under the combination of time-reversal and spatial translation operations. 
When introducing $H$ in the $z$ direction, this symmetry is broken, and the $z$ component of $\bar{\mathbf{b}}$, $\bar{b}_z$, becomes nonzero and takes a negative value, while the $x$ and $y$ components remain zero, as shown in the middle panel of Fig.~\ref{fig:gs_hdep}. 
The magnitude of $\bar{b}_z$ increases with $H$ in the $\Nt=4$ phase and decreases in the $\Nt=2$ phase, leaving a peak at the topological transition. 
Since the Magnus force originates from the total EMF, transverse motions of torons can be caused at nonzero magnetic fields (see Figs.~3\textbf{a}--\textbf{c} in the main text)~\cite{Jonietz2010,Zang2011,Iwasaki2013,Jiang2017}. 
While increasing $H$, each hedgehog and antihedgehog moves towards its counterpart along the skyrmion string. 
This leads to the shrinkage of skyrmion strings, as shown in the bottom panel of Fig.~\figref{gs_hdep}{}: 
Two out of four torons reduce the lengths of skyrmion strings, $\dt$, from $7$ to $6$, and the rest two from $5$ to $4$. 
The latter exhibits a rapid drop and becomes zero toward the topological transition at $H\simeq H_{\mathrm{topo}}$, corresponding to pair annihilation of the hedgehogs and antihedgehogs. 
Meanwhile, the former remains unchanged until the rapid decrease before the phase transition to the $2Q$ phase. 
These changes of $\dt$ are summarized in Fig.~2 in the main text.

\begin{figure*}[tb]
\centering
\includegraphics[width=1.0\columnwidth]{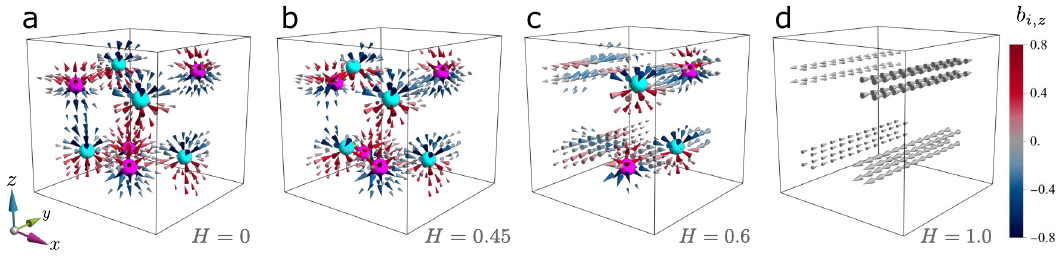}
\caption{
\label{fig:EMF}
\textbf{Emergent magnetic field in the $3Q$-HL.}
\textbf{a},\textbf{b},\textbf{c},\textbf{d}  
Real-space distribution of the emergent magnetic field $\mathbf{b}_i$ for the $3Q$-HL within the MUC at $H=0$ (\textbf{a}), $H=0.45$ (\textbf{b}), $H=0.6$ (\textbf{c}), and $H=1.0$ (\textbf{d}). 
The arrows represent $\mathbf{b}_i$ and their color displays the $z$ component $b_{i,z}$ (see the inset in \textbf{d}). 
Only $\mathbf{b}_i$ with large amplitude are shown and the divergently large lengths close to the hedgehogs and antihedgehogs are normalized for clarity. 
}
\end{figure*}

Figure~\figref{EMF}{} displays the spatial distributions of the Bloch points and the local EMF at site $i$, $\mathbf{b}_i$. 
At $H=0$, although the average EMF $\bar{\mathbf{b}}$ is zero, $\mathbf{b}_i$ is locally nonzero and has a large amplitude around the Bloch points due to the strong twist of spins, as shown in Fig.~\figref{EMF}{a}. 
The EMF emanates from the hedgehogs (emergent magnetic monopoles) and flows into the adjacent antihedgehogs (emergent magnetic antimonopoles); see also Fig.~1\textbf{b} in the main text. 
When introducing $H$, the Bloch points change their positions, and the associated modulation of $\mathbf{b}_i$ makes the average EMF $\bar{b}_z$ nonzero. 
The hedgehog and antihedgehog in each toron approach each other by reducing $\dt$, thus amplifying the EMF between them monotonically in the $N_{\mathrm{t}}=4$ phase. Figure~\figref{EMF}{b} shows such a spatial distribution at $H=0.45$, where $\dt=2$ for two out of four torons and $\dt=6$ for the rest two; $\mathbf{b}_i$ becomes stronger especially around the torons with $\dt=2$. 
By further increasing $H$, the $3Q$-HL exhibits the topological transition to the $N_{\mathrm{t}}=2$ phase. 
The result at $H=0.6$ presented in Fig.~\figref{EMF}{c} shows that $\mathbf{b}_i$ appears around not only the remaining torons but also the positions where the torons annihilate at $H \simeq H_{\mathrm{topo}}$. 
In this phase, $-\bar{b}_z$ decreases by increasing $H$. 
We also show the result for the $2Q$ state at $H=1.0$ in Fig.~\figref{EMF}{d}. 
In this case also, residual EMFs persist near the positions where the four torons vanish. 
Since $\SSF{\mathbf{Q}_2}=0$, there is no spin modulation in the $y$ direction, and thus $\mathbf{b}_i$ aligns parallel to the $y$ axis; however, $b_{i,y}$ appears in an alternating manner, leading to $\bar{\mathbf{b}}=0$.

\section{Origin of the vanishingly small potential barrier}

\begin{figure*}[tb]
\centering
\includegraphics[width=1.0\columnwidth]{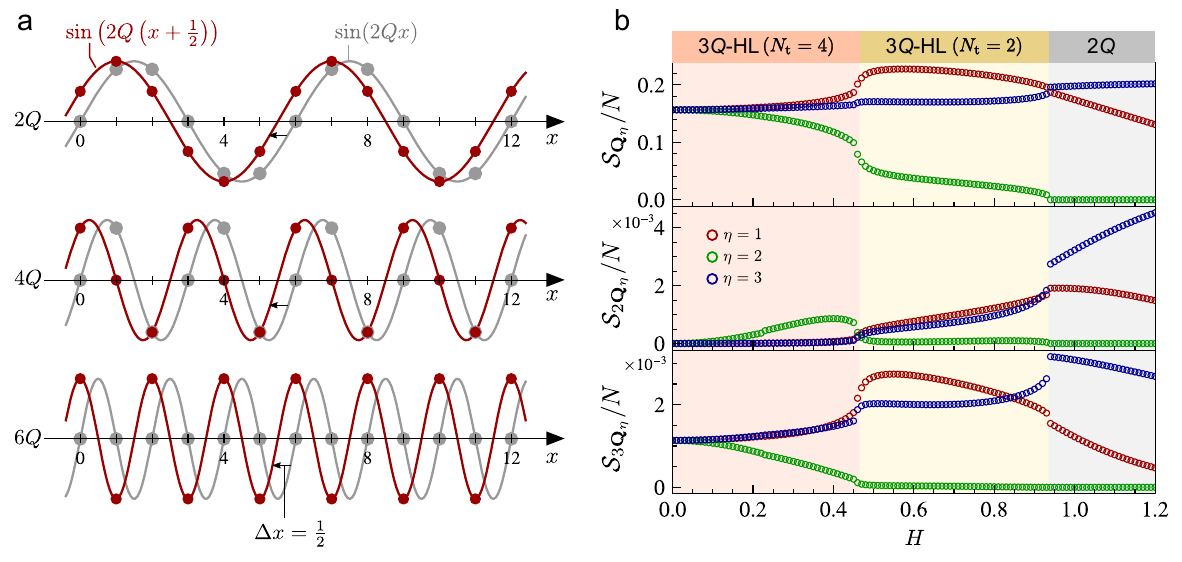}
\caption{
\label{fig:SSF_harmonics}
\textbf{Higher harmonics of the spin structure factor and their effect from the lattice discretization}. 
\textbf{a} Schematic figures for the spatial translation by a half lattice constant $\Delta x=\frac12$ for the higher harmonics. The lines and dots represent the higher harmonics in continuous space and their values on the lattice sites, respectively. The original higher harmonics and those after the translation are denoted by the gray and red colors, respectively. Presented here are the higher harmonics with the wave numbers of $2Q$ (top), $4Q$ (middle), and $6Q$ (bottom) with $Q=\frac{\pi}{6}$. 
\textbf{b} $H$ dependences of $\mathcal{S}_{\mathbf{Q}_{\eta}}$ (top), $\mathcal{S}_{2\mathbf{Q}_{\eta}}$ (middle), and $\mathcal{S}_{3\mathbf{Q}_{\eta}}$ (bottom). 
}
\end{figure*}

In Fig.~4\textbf{c} in the main text, we find that the potential barrier in the $y$ direction, $V_y$, becomes vanishingly small in the $3Q$-HL phase with $N_{\mathrm{t}}=2$. 
This appears to be attributed to strong suppression of the higher harmonics of the spin structure factor beyond the Nyquist wave number as follows. 

In general, a superposition of multiple spin density waves exhibits higher harmonics, such as $\mathcal{S}_{2{\bf Q}_1}$ and $\mathcal{S}_{{\bf Q}_1+{\bf Q}_2}$, through the normalization of the spin length, $|\mathbf{S}_i|=1$. 
The higher harmonics represent rapidly oscillating components in real space, and hence they can be largely affected by the lattice discretization. 
This is schematically shown in Fig.~\figref{SSF_harmonics}{a}.
When the wave number is smaller than $6Q=\pi$, the higher harmonics restored from the values on the lattice sites before and after the translation by half a lattice constant are identical (the upper two panels of Fig.~\figref{SSF_harmonics}{a}). 
However, the original wave cannot be restored for the case with $6Q$; the inverse Fourier transform of the gray dots taking zero cannot reproduce the gray lines, while the red dots can do (the lowest panel of Fig.~\figref{SSF_harmonics}{a}). 
This wave number corresponds to the Nyquist wave number. 
Thus, when the spin state includes such higher harmonics equal to or greater than the Nyquist wave number, its spatial translation costs energy and results in a potential barrier. 

Figure~\figref{SSF_harmonics}{b} shows the $H$ dependences of a series of $\mathcal{S}_{n{\bf Q}_{\eta}}$ ($n=1,2,3$) in the present model. 
In the $3Q$ and $2Q$ states, the higher harmonics $\SSF{2\mathbf{Q}_{\eta}}$ and $\SSF{3\mathbf{Q}_{\eta}}$ become nonzero for all the components $\eta$ included in the superpositions, as expected. 
However, in the $3Q$-HL with $N_{\mathrm{t}}=2$ for $0.465 \lesssim H \lesssim 0.935$, $\SSF{2\mathbf{Q}_{2}}$ is strongly suppressed and $\SSF{3\mathbf{Q}_{2}}$ becomes vanishingly small. 
We confirm that further higher harmonics $\SSF{n\mathbf{Q}_{\eta}}$ ($n\geq4$) are also negligibly small only for $\eta=2$ (not shown). 
This means that the $3Q$ state hardly feels the lattice discretization for a translation in the $y$ direction. 
This is the reason why the potential barrier $V_y$ is vanishingly small in the $3Q$-HL phase with $N_{\mathrm{t}}=2$.

\section{Current-velocity relation for $\tilde{\mathbf{j}}_e \parallel \hatv{z}$ \label{sec:j||z}}

\begin{figure*}[tb]
\centering
\includegraphics[width=0.6\columnwidth]{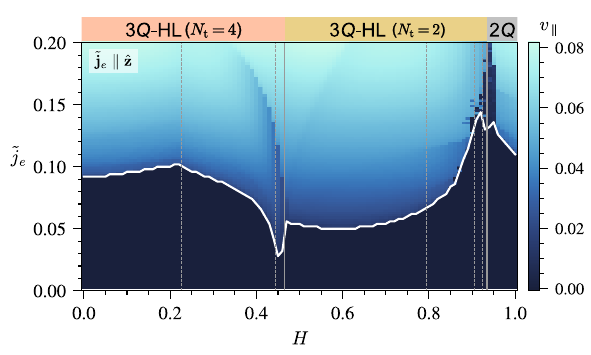}
\caption{
\label{fig:current_velocity_z}
\textbf{Current-velocity relation for $\tilde{\mathbf{j}}_e \parallel \hatv{z}$.} 
The velocity of the spin texture for the longitudinal motions, $v_{\parallel}$, with $\tilde{\mathbf{j}}_e\parallel\hatv{z}$. 
The notations are common to those in Fig.~3\textbf{a} in the main text. 
}
\end{figure*}

We discuss the drift motions of torons caused by an electron flow $\tilde{\mathbf{j}}_e$ in the $z$ direction. 
In this case, the toron Hall effect does not occur since $\tilde{\mathbf{j}}_e \parallel \bar{\mathbf{b}}$, while the longitudinal drift motion does. 
The threshold current density $\tilde{j}_{e,\mathrm{c}\parallel}$ and the velocity $v_{\parallel}$ for the longitudinal motion exhibit different $H$ dependences from those for $\tilde{\mathbf{j}}_e \parallel \hatv{x}$ and $\hatv{y}$ respectively shown in Figs.~3\textbf{a} and 3\textbf{d} in the main text. 
The results are presented in Fig.~\figref{current_velocity_z}{}. 
By increasing $H$ from zero, the threshold current $\tilde{j}_{e,\mathrm{c}\parallel}$ increases slowly and shows a cusp at the discontinuous change of $d_{\mathrm{t}}$ at $H \simeq 0.225$. 
A similar $H$ dependence is observed in the $\tilde{\mathbf{j}}_e \parallel \hatv{x}$ case shown in Fig.~3\textbf{a}, where an electron flow is perpendicular to the skyrmion strings. 
By further increasing $H$, however, $\tilde{j}_{e,\mathrm{c}\parallel}$ turns to decrease, exhibiting a dip near the topological transition at $H \simeq H_{\mathrm{topo}}$. 
This trend is similar to the $\tilde{\mathbf{j}}_e \parallel \hatv{y}$ case shown in Fig.~3\textbf{d}, where the electron flow is parallel to the skyrmion strings. 
After the topological transition at $H=H_{\mathrm{topo}} \simeq 0.465$, $\tilde{j}_{e,\mathrm{c}\parallel}$ exhibits a rapid increase with $H$ toward the phase transition to the $2Q$ state at $H = H_{\mathrm{c}} \simeq 0.935$, again similar to the $\tilde{\mathbf{j}}_e \parallel \hatv{x}$ case in Fig.~3\textbf{a}. 
The $H$ dependence of $\tilde{j}_{e,\mathrm{c}\parallel}$ is well explained from that of $V_z$ shown in Fig.~4\textbf{c} in the main text, similar to the cases with $\tilde{\mathbf{j}}_e \parallel \hatv{x}$ and $\tilde{\mathbf{j}}_e \parallel \hatv{y}$. 

These trends of $\tilde{j}_{e,\mathrm{c}\parallel}$ are qualitatively understood by the spatial structures of the skyrmion strings (see Figs.~2\textbf{d}--\textbf{g} in the main text). 
In the weak field region for $H\lesssim 0.225$, the skyrmion strings extend predominantly in the $y$ direction. 
Consequently, these strings, which are perpendicular to the current $\tilde{\mathbf{j}}_e \parallel \hatv{z}$, primarily contribute to $\tilde{j}_{e,\mathrm{c}\parallel}$, thereby giving rise to the similar trends to $\tilde{\mathbf{j}}_e \parallel \hatv{x}$. 
For $H \gtrsim 0.225$, however, the skyrmion strings reduce their length in the $y$ direction, relatively enhancing the effect from the kink structures in the skyrmion strings. 
At the kinks, the skyrmion strings run in the $z$ direction, bringing about the similar trend to $\tilde{\mathbf{j}}_e \parallel \hatv{y}$. 
After the topological transition, the kink structures disappear in the $3Q$-HL with $N_{\mathrm{t}}=2$, resulting in the similar trend to $\tilde{\mathbf{j}}_e \parallel \hatv{x}$ again.

\section{Parameters in the superposition of three elliptical spirals \label{sec:parameter}}

\begin{figure*}[tb]
\centering
\includegraphics[width=0.6\columnwidth]{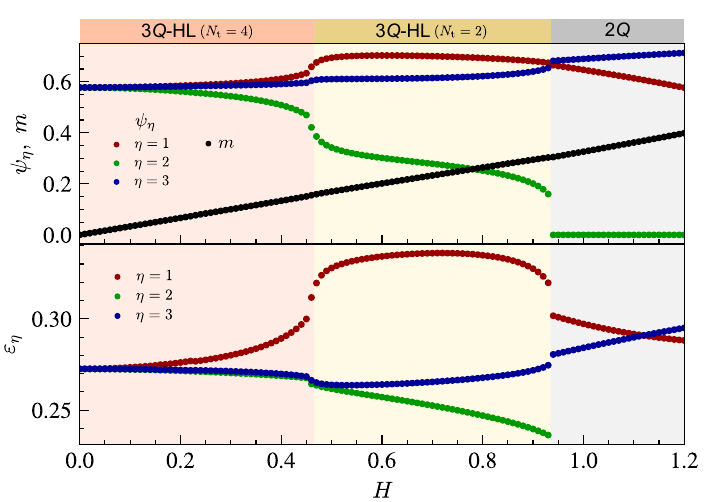}
\caption{
\label{fig:fit_params_hdep}
\textbf{Parameters in the superposition of three elliptical spirals.} 
$H$ dependence of the amplitude $\psi_{\eta}$ and the uniform magnetization $m$ (top), and the ellipticity $\varepsilon_{\eta}$ (bottom) in the superposition of three elliptical spirals given by \refEqSuperposition{} in the main text. 
The results are obtained by evaluating the loss function given by \refEqLoss{} in the main text. 
}
\end{figure*}

We show in Fig.~\figref{fit_params_hdep}{} the $H$ dependences of the parameters in the superposition of the three elliptical spirals [\refEqSuperposition{} in the main text] estimated by minimizing the loss function [\refEqLoss{} in the main text]. 
As shown in the upper panel, the amplitudes of the spirals, $\psi_{\eta}$, behave similarly to $\mathcal{S}_{\mathbf{Q}_{\eta}}$ in the top panel of Fig.~\figref{gs_hdep}{}. 
The three $\psi_{\eta}$ are identical at $H=0$, while all of them become inequivalent as $\psi_1 > \psi_3 > \psi_2 >0$. 
In the $2Q$ phase, $\psi_2$ becomes zero, and thus the spin texture is represented by the superposition of two spirals. 
The uniform magnetization $m$ increases almost linearly with $H$. 
In the bottom panel, we show the evolution of the ellipticity for each spiral, $\varepsilon_{\eta}$. 
Although the Dzyaloshinskii-Moriya (DM) interaction prefers circular spirals with $\varepsilon_{\eta} = 1$, the exchange anisotropy $\Delta$ in \refEqModel{} in the main text deforms them into ellipses, resulting in  $0 < \varepsilon_{\eta} < 1$. 
Similar to $\psi_{\eta}$, $\varepsilon_{\eta}$ are identical among three spirals at $H=0$ and become inequivalent by introducing $H$ as $0 < \varepsilon_2 \leq \varepsilon_3 \leq \varepsilon_1 <1$.

\section{Results without the DM interaction}
\label{sec:D=0}

The results discussed in the main text are obtained for the model in \refEqModel{} in the presence of the DM interaction with $D=0.3$. 
However, the DM interaction is not always necessary to stabilize the HL~\cite{Okumura2022}. 
In this section, we show the results with $D=0$ for comparison.

\subsection{Ground-state property \label{sec:gs_D=0}}

\begin{figure*}[tb]
\centering
\includegraphics[width=1.0\columnwidth]{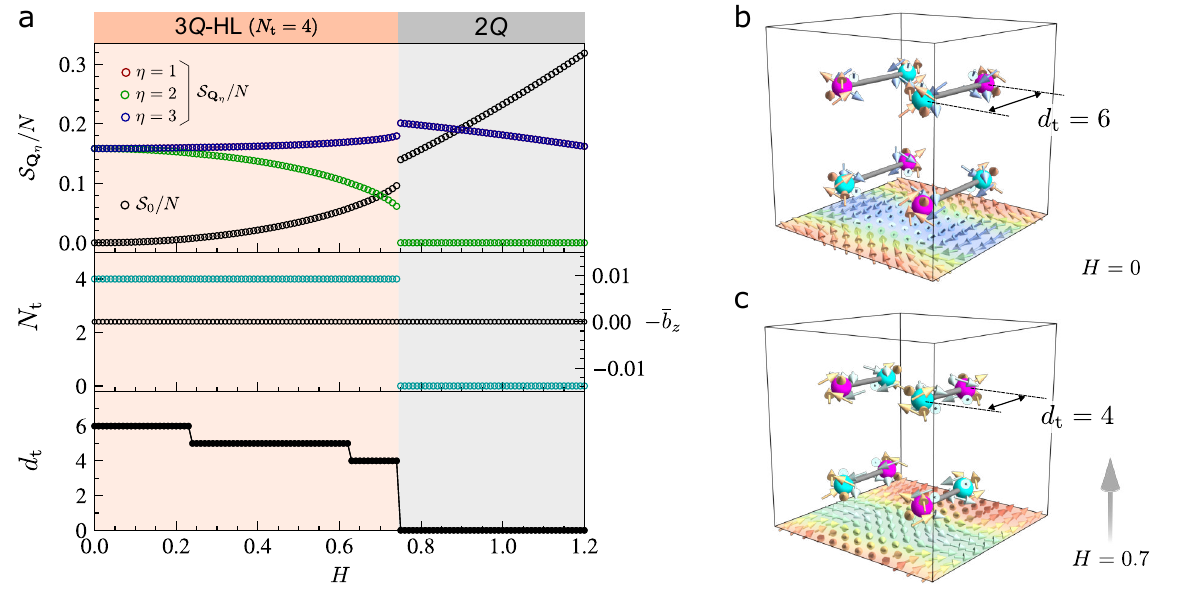}
\caption{
\label{fig:gs_hdep_D=0}
\textbf{Magnetic and topological properties in the ground state in the absence of the DM interaction.} 
\textbf{a} Similar figure to Fig.~\ref{fig:gs_hdep} for the model in \refEqModel{} in the main text with $D = 0$. 
\textbf{b},\textbf{c} Real-space distribution of the hedgehogs and antihedgehogs, the skyrmion strings, and spins around the hedgehogs and antihedgehogs within the MUC for the ground state at $H=0$ (\textbf{b}) and $H=0.7$ (\textbf{c}). 
The notations are common to those in Figs.~2\textbf{d}--\textbf{g} in the main text. 
}
\end{figure*}

Figure~\figref{gs_hdep_D=0}{a} summarizes the ground-state properties for the model in \refEqModel{} in the main text with $D=0$. 
The other parameters are the same as in the main text. 
See Sec.~\ref{sec:gs} for the definitions of the physical quantities. 
The spin structure factor $\SSFq$ displayed in the top panel shows that the system is in a $3Q$ state for $0 \leq H \lesssim 0.745$, and turns into a $2Q$ state for $H \gtrsim 0.745$. 
In the $3Q$ state, three $\SSF{\mathbf{Q}_{\eta}}$ are identical at zero field and $\SSF{\mathbf{Q}_1}=\SSF{\mathbf{Q}_3} > \SSF{\mathbf{Q}_2} > 0$ for nonzero fields. 
Note that, in the present case with $D=0$, the Hamiltonian in \refEqModel{} is invariant by simultaneously exchanging $x$ and $z$ coordinates, and $S_x$ and $S_y$, and hence, $\SSF{\mathbf{Q}_1}$ is always identical to $\SSF{\mathbf{Q}_3}$. 
At $H \simeq 0.745$, the $3Q$ state turns into the $2Q$ state accompanied by the discontinous changes in $\SSF{\mathbf{Q}_{\eta}}$ as well as $\SSF{0}$. 
In the $2Q$ state, $\SSF{\mathbf{Q}_2}=0$ and $\SSF{\mathbf{Q}_1}=\SSF{\mathbf{Q}_3} > 0$. 
As shown in the middle panel, $N_{\mathrm{t}}$ is $4$ in the $3Q$ state, and hence this state is the topologically nontrivial $3Q$-HL with four torons in the MUC. 
Meanwhile, in contrast to the results in Fig.~\figref{gs_hdep}{}, the $3Q$-HL with $N_{\mathrm{t}}=4$ directly turns into the $2Q$ state without showing the topological transition by the annihilation of half of the torons. 
In this $D=0$ case, $\bar{b}_z$ is always zero in the entire field region.

The real-space distribution of the hedgehogs and antihedgehogs, and the skyrmion strings at $H=0$ is shown in Fig.~\figref{gs_hdep_D=0}{b}. 
The hedgehogs and antihedgehogs form the NaCl structure and they are paired by the skyrmion strings running in the $y$ direction with the length of $d_{\mathrm{t}}=6$. 
By increasing $H$, the hedgehogs and antihedgehogs move toward each other by shortening the skyrmion strings, as shown in Fig.~\figref{gs_hdep_D=0}{c} at $H=0.7$. 
We show the $H$ dependence of $d_{\mathrm{t}}$ in the bottom panel of Fig.~\figref{gs_hdep_D=0}{a}.  
In this case with $D=0$, $\dt$ for the four torons change in the same way; they decrease from 6 to 4 successively by increasing $H$, but suddenly become zero at the phase transition to the $2Q$ state. 
We note that the skyrmion strings are always straight without kink structures, in contrast to the case with $D=0.3$. 
Such an evolution keeps the high-symmetric distribution of hedgehogs and antihedgehogs, leading to $\bar{b}_z=0$, while the local EMF is large around the hedgehogs and antihedgehogs similar to Fig.~\figref{EMF}{}.

\subsection{Current-velocity relation \label{sec:current_velocity_D=0}}

\begin{figure*}[tb]
\centering
\includegraphics[width=1.0\columnwidth]{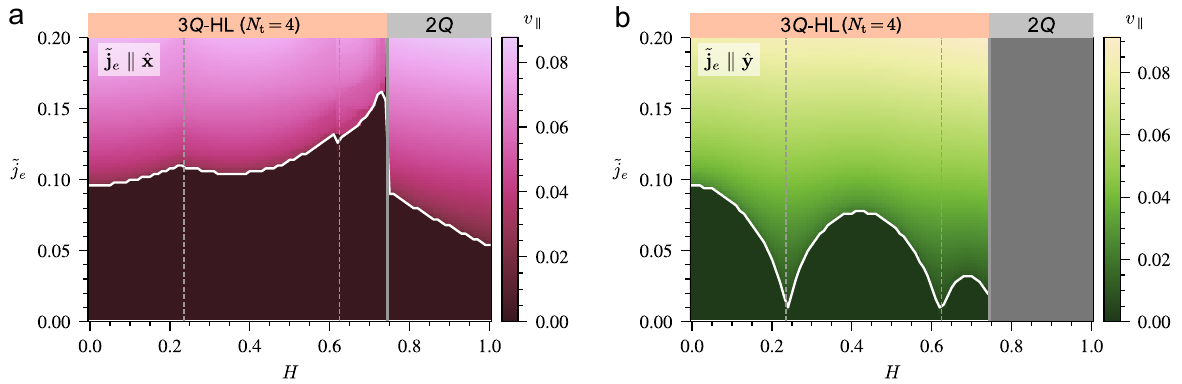}
\caption{
\label{fig:current_velocity_D=0}
\textbf{Current-velocity relations in the absence of the DM interaction.} 
\textbf{a},\textbf{b} The velocities for the longitudinal motions with $\tilde{\mathbf{j}}_e\parallel\hatv{x}$ (\textbf{a}) and $\tilde{\mathbf{j}}_e\parallel\hatv{y}$ (\textbf{b}) for the model in \refEqModel{} in the main text with $D=0$. 
The result for the longitudinal velocity with $\tilde{\mathbf{j}}_e\parallel\hatv{z}$ is common to \textbf{a} by symmetry (see Sec.~\ref{sec:gs_D=0} for the details). 
}
\end{figure*}

In the model in \refEqModel{} in the main text with $D=0$, the drift motion occurs only along the current direction and the toron Hall effect does not occur due to the vanishing average EMF $\bar{\mathbf{b}}=0$. 
Figure~\figref{current_velocity_D=0}{} displays the threshold current density $\tilde{j}_{e,\mathrm{c}\parallel}$ and the velocity $v_{\parallel}$ for the longitudinal motions. 
In the case with $\tilde{\mathbf{j}}_e \parallel \hatv{x}$ (Fig.~\figref{current_velocity_D=0}{a}), the longitudinal drift motion occurs above $\tilde{j}_{e,\mathrm{c}\parallel}$ represented by the white line, and $v_{\parallel}$ monotonically increases with $\tilde{j}_{e}$ above the threshold.  
By increasing $H$, $\tilde{j}_{e,\mathrm{c}\parallel}$ overall increases while showing cusps at $H\simeq0.235$ and $H\simeq0.625$, corresponding to the changes of $d_{\mathrm{t}}$ from 6 to 5 and from $5$ to $4$, respectively (the bottom panel of Fig.~\figref{gs_hdep_D=0}{a}). 
Similar cusp structures in the $H$ dependence of $\tilde{j}_{e,\mathrm{c}\parallel}$ are also observed in Fig.~3\textbf{a} in the main text. 
$\tilde{j}_{e,\mathrm{c}\parallel}$ exhibits a sudden drop at the phase transition to the $2Q$ state at $H\simeq0.745$ denoted by the solid gray line, and decreases by increasing $H$. 
We note that essentially the same results are obtained for $\tilde{\mathbf{j}}_e \parallel \hatv{z}$ (not shown) due to the symmetry discussed in Sec.~\ref{sec:gs_D=0}.

Figure~\figref{current_velocity_D=0}{b} shows the results for $\tilde{\mathbf{j}}_e \parallel \hatv{y}$. 
Similar to Fig.~3\textbf{d} in the main text, $\tilde{j}_{e,\mathrm{c}\parallel}$ exhibits the dips corresponding to the shrinkage of the skyrmion strings. 
The maximum value of $\tilde{j}_{e,\mathrm{c}\parallel}$ in each dome decreases with the reduction of $d_{\mathrm{t}}$, also similar to Fig.~3\textbf{d} in the main text. 
Note that $v_{\parallel}$ cannot be defined in the $2Q$ state since the spin modulation in the $y$ direction, $\SSF{\mathbf{Q}_2}$, vanishes.

\subsection{Potential barrier \label{sec:potential_D=0}}

\begin{figure*}[tb]
\centering
\includegraphics[width=0.5\columnwidth]{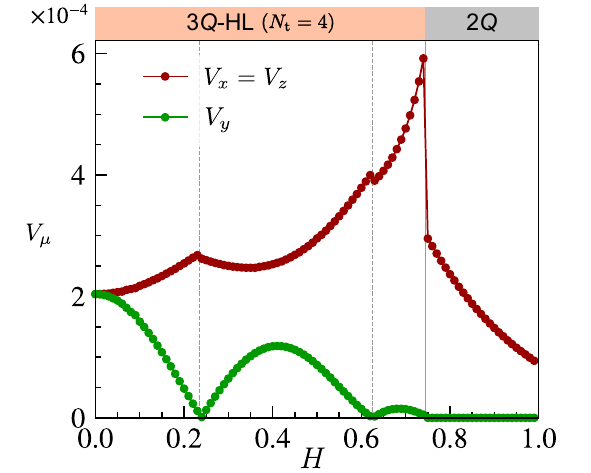}
\caption{
\label{fig:potential_hdep_D=0}
\textbf{Potential barrier for monopoles in the absence of the DM interaction.}
$H$ dependence of the potential barriers $V_{\mu}$ for the model in \refEqModel{} in the main text with $D=0$. 
See Methods in the main text for the details of the calculation. 
}
\end{figure*}

We calculate the potential barrier for torons by following the case with $D=0.3$ (see Methods in the main text). 
First, we obtain the optimal parameters in the superposition of the three spin density waves given by \refEqSuperposition{} in the main text by minimizing the loss function \refEqLoss{} in the main text. 
In the present case with $D=0$, we find that the absence of the DM interaction and the presence of the exchange anisotropy lead to $\varepsilon_{\eta}=0$, and hence the ground-state spin configuration is represented by a superposition of sinusoidal waves. 
Then, we calculate the energy change per site for the optimal spin configuration by shifting the phases $\vp_{\eta}$ and obtain the potential barrier in the $\mu$ direction, $V_{\mu}$. 

Figure~\figref{potential_hdep_D=0}{} shows the $H$ dependences of $V_{\mu}$ for $D=0$. 
Similar to Fig.~4\textbf{c} in the main text, $V_x=V_y=V_z$ at zero field due to symmetry. 
By increasing $H$, $V_x$ and $V_z$ increase keeping the relation $V_x=V_z$, and show the cusps at the discontinuous changes of $d_{\mathrm{t}}$ at $H\simeq0.235$ and $H\simeq0.625$. 
They increase further toward the phase transition to the $2Q$ state at $H\simeq0.745$, but decrease in the $2Q$ state after the sharp drop at the phase transition. 
In contrast, $V_y$ decreases with increasing $H$ and shows the dips at the discontinuous changes of $d_{\mathrm{t}}$ at $H\simeq0.235$ and $H\simeq0.625$, and the peak value of $V_y$ in each dome decreases toward the phase transition. 
These $H$ dependences well explain those of $\tilde{j}_{e,\mathrm{c}\parallel}$ shown in Fig.~\figref{current_velocity_D=0}{}.

\section*{Description of the supplementary videos}

\begin{description}
\item[Supplementary Video 1]
\textbf{Perfect toron Hall effect.} 
Real-space and real-time distributions of the magnetic torons within the MUC for the $3Q$-HL ($N_{\mathrm{t}}=4$) at $H=0.3$ with $\tilde{\mathbf{j}}_e\parallel\hatv{x}$ and $\tilde{j}_e=0.1$. The torons exhibit the transverse motion without any longitudinal one. The notations are common to those in Figs.~2 \textbf{d}--\textbf{g} in the main text. 
\item[Supplementary Video 2]
\textbf{Zero toron Hall effect.} 
Similar video to Supplementary Video 1 at $H=0.3$ with $\tilde{\mathbf{j}}_e\parallel\hatv{x}$ and $\tilde{j}_e=0.16$. 
The torons exhibit the longitudinal motion without any transverse one. 
\item[Supplementary Video 3]
\textbf{Toron Hall effect.} 
Similar video to Supplementary Video 1 at $H=0.3$ with $\tilde{\mathbf{j}}_e\parallel\hatv{x}$ and $\tilde{j}_e=0.2$. 
The torons exhibit both longitudinal and transverse motions. 
\end{description}

\bibliography{ref}